\documentclass[12pt,preprint]{aastex} 

\slugcomment{Received 2009 May 29; accepted 2009 June 30}

\def\calg{{\cal G}}

\def\esn{E_{\rm SN}}

\def\rp{r_{\rm p}} 
\def\rsigma{r_{\sigma}} 
\def\Rsigma{R_{\sigma}}

\def\taup{\tau_{\rm p}} 
\def\tp{t^{\prime}} 
 
\def\vr{V_{\rm R}}

\def\siggas{\Sigma}
\def\sigsfr{\dot{\Sigma}_{\star}} 
 
\def\sunm{M_{\odot}}

\def\ergs{\ifmmode {\rm ergs~ s^{-1}} \else {\rm ergs~s^{-1}}\ \fi} 
\def\kms{\ifmmode {\rm km~ s^{-1}} \else {\rm km~s^{-1}}\ \fi} 
 
\def\mbh{M_{\bullet}} 
 
\def\mgii{\ifmmode Mg {\sc ii} \else Mg {\sc ii}\ \fi}

\def\sunm{M_{\odot}}

\def\ergs{${\rm erg~s^{-1}}$}

\begin{document}

\title{Evolution of gaseous disk viscosity driven by supernova explosion in star-forming  
galaxies at high redshift}

\author{
Jian-Min Wang\altaffilmark{1,2},  
Chang-Shuo Yan\altaffilmark{1,3}, 
Yan-Rong Li\altaffilmark{1,3},  
Yan-Mei Chen\altaffilmark{1,3}, 
Fei Xiang\altaffilmark{1}, \\ 
Chen Hu\altaffilmark{1}, 
Jun-Qiang Ge\altaffilmark{1,3}, and  
Shu Zhang\altaffilmark{1} 
} 
 
\altaffiltext{1} 
{Key Laboratory for Particle Astrophysics, Institute of High Energy Physics, 
Chinese Academy of Sciences, 19B Yuquan Road, Beijing 100049, China}

\altaffiltext{2} 
{Theoretical Physics Center for Science Facilities, Chinese Academy of Sciences, China} 

\altaffiltext{3} 
{Graduate University of Chinese Academy of Sciences, 19A Yuquan Road, Beijing 100049, China}

\begin{abstract} 
Motivated by Genzel et al.'s observations of high-redshift star-forming galaxies, containing 
clumpy and turbulent rings or disks, we build a set of equations describing the dynamical 
evolution of gaseous disks with inclusion of star formation and its feedback. Transport of 
angular momentum is due to "turbulent" viscosity induced by supernova explosions in 
the star formation region. Analytical solutions of the equations are found for the initial 
cases of a gaseous ring and the integrated form for a gaseous disk, respectively. For a  
ring with enough low viscosity, it evolves in a slow processes of gaseous diffusion and star 
formation near the initial radius. For a high viscosity, the ring rapidly diffuses in the 
early phase. The diffusion drives the ring into a region with a low viscosity and start the 
second phase undergoing pile-up of gas at a radius following the decreased viscosity torque. 
The third is a sharply deceasing phase because of star formation consumption of gas and 
efficient transportation of gas inward forming a stellar disk. We apply the model to two 
$z\sim 2$ galaxies BX 482 and BzK 6004, and find that they are undergoing a decline in their 
star formation activity.
\end{abstract} 
 
\section{Introduction} 
Exciting progress has been made in observing assembly of gas and star formation in galaxies from 
low to high redshifts. Optically/UV-selected star-forming galaxies at $z\sim 0.5--2.5$ exhibit 
a fairly tight relationship between stellar mass and star formation rate, which is strongly 
suggestive of star formation happening with a high duty cycle (Noeske et al. 2007; Elbaz et al. 
2007; Daddi et al. 2007). This may relate with random accretion onto supermassive black holes
(SMBHs) triggered through minor mergers located at galactic centers (Wang et al 2009), showing 
episodic activities of the SMBHs (Wang et al. 2006, 2008). Major mergers are undoubtedly taking 
place in very luminous submillimeter 
galaxies at $z\sim 1--3$ (Tacconi et al. 2006; 2008), but only a small fraction of this sample 
($\sim 1/3$) is undergoing obvious major mergers from the images of the sample (F\"oster Schreiber 
et al. 2006; Genzel et al. 2006; Wright 2007; Law et al. 2007).  Detail studies of 
Spectroscopic Imaging survey in the Near-infrared with SINFONI (SINS) of high-redshift galaxies 
show that about 1/3 of the samples are rotation-dominated turbulent star-forming rings/disks; another 
one-third are compact and velocity dispersion-dominated objects and the remains are clear 
interacting/merging systems (Genzel et al. 2008; F\"oster Schreiber et al. 2009). These dynamical 
evidences show that rapid and continuous gas accretion via "cold flows" and/or minor mergers are 
playing an important role in driving star formation and assembly of star-forming galaxies at $z>1$ 
(Genzel et al. 2008). Interestingly, recent numerical simulations produce results agreeing with 
the observations (Bower et al. 2006; Kitzbichler \& White 2007; Naab et al. 2007; Guo \& White 2008; 
Dav\'e 2008;  Brooks et al. 2009; Dekel et al. 2009). Minor mergers as a supply of gas may trigger 
star formation and subsequently ignite the SMBHs. The role of supernova explosions (SNexp) in driving 
interstellar turbulence and the existence of a supernova rate--velocity dispersion relation has 
been highlighted using numerical simulations (e.g., Wada \& Norman 2002; Dib et al. 2006), and 
observationally by the spatial correlations of the inclinations of molecular clouds with respect 
to the Galactic disk attributed to the effects of SNexp (Dib et al. 2009; see an extensive review 
of Mac Low \& Klessen 2004), and by the existence of a starburst-AGN connection (Chen et al. 2009). 
SNexp is potentially playing a role in the dynamics in star-forming galaxies.

Distributions of density and angular momentum of a gaseous disk are generally controlled by the 
viscosity, if driven by viscosity of  turbulence excited by SNexp, stellar distribution will be 
determined by this feedback process (Lin \& Pringle 1987). Steady analytical solution of 
circumnuclear disk of the Galaxy with injection of SNexp energy has been obtained by Vollmer \& 
Beckert (2003), coevolution of SMBHs and their hosts could be driven by 
SNexp-excited turbulence (Kawakatu \& Wada 2008) as 
well as in the starburst--AGN connection in light of strong correlation between the Eddington ratios 
and specific star formation rates (Chen et al. 2009). Figure 1 shows a clear trend of concentration of 
galaxies with star formation rate densities in the SINS sample given by Genzel et al. (2008), 
strengthening the interplay and regulation between SN explosions, cold gas accretion, 
and minor mergers.

In this Letter, we consider the secular evolution of gaseous disk, which could be formed during 
minor mergers, by including feedback from SNexp in star formation or burst regions characterized 
by the H$\alpha$ ring in the SINS sample. We find the analytical solutions, which show two kinds 
of properties: 1) piled-up gas and 2) fast diffusion depending on the initial conditions of gas 
density. We apply the models to high-redshift galaxies.

{\centering 
\figurenum{1} 
\includegraphics[angle=-90,scale=0.25]{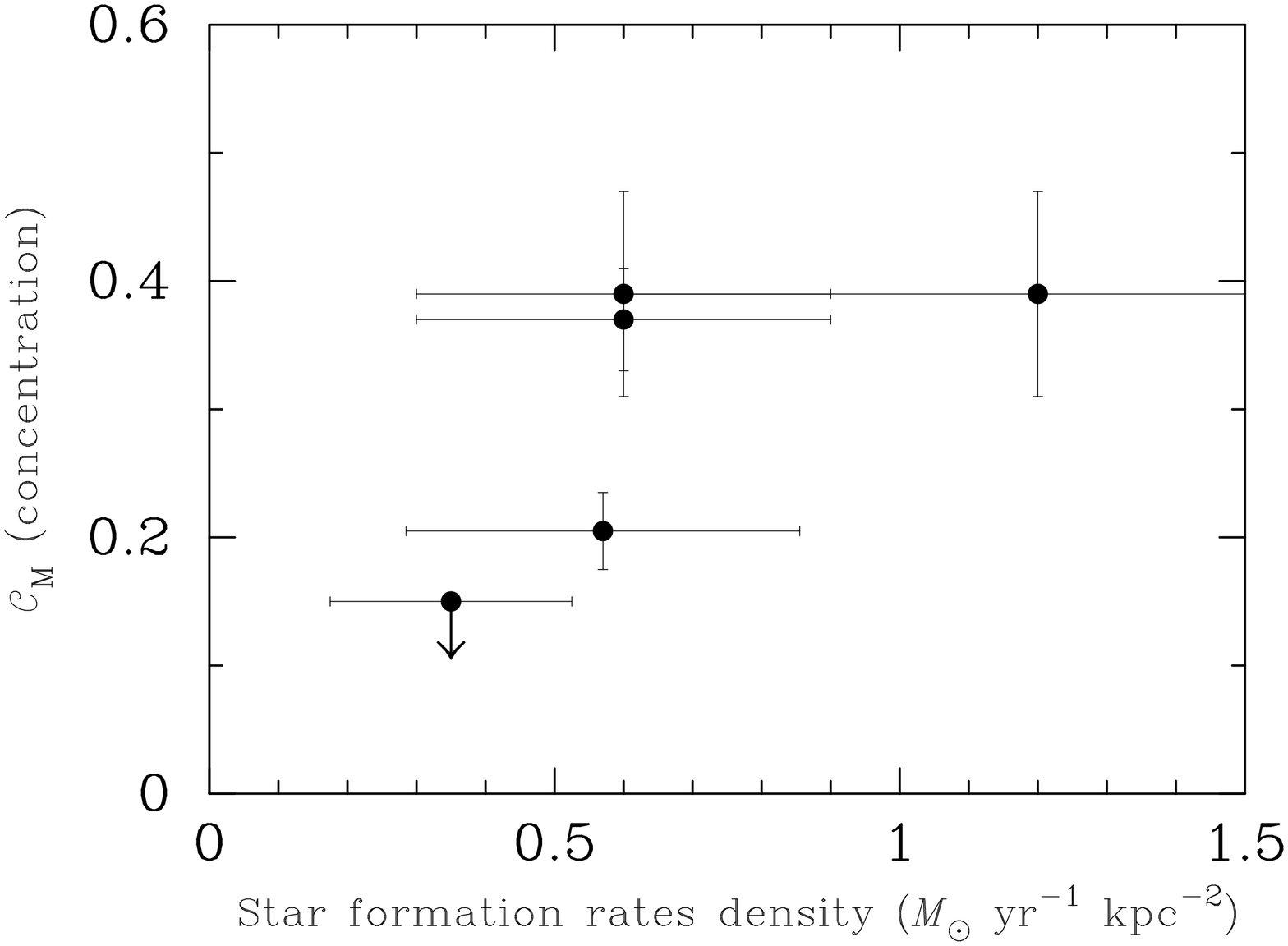} 
\figcaption{\footnotesize Mass concentration of galaxies defined by a ratio of dynamical mass 
within $0^{\prime\prime}.4$-$1^{\prime\prime}.2$, 
${\cal C}_{\rm M}=M_{\rm dyn}(\le 0^{\prime\prime}.4)/M_{\rm dyn}(\le 1^{\prime\prime}.2)$, 
correlates with star formation rates for the SINS sample given in Table 2 in Genzel et al. (2008). 
This correlation strengthens the role of SNexp. 
} 
\label{fig1} 
\vglue 0.3cm}

\section{Dynamical equations} 

Considering the mass dropout of gaseous disk due to ongoing star formation, the mass conservation 
equation reads
%
\begin{equation} 
R\frac{\partial\siggas}{\partial t} 
+\frac{\partial}{\partial R}\left(R\vr\siggas\right) +R\sigsfr=0, 
\end{equation} 
where $\vr$ is the radial velocity of the gas, $\siggas$ is the surface density of gas, and $\sigsfr$ 
is the surface density of star formation rates. We neglect mass injection of stellar winds back to 
the gaseous disk or escapes of winds from galaxies. Star formation process removes some angular 
momentum of gas, thus the conservation equation of angular momentum is given by  
\begin{equation} 
R\frac{\partial}{\partial t}\left(\siggas R^2\Omega\right)+ 
\frac{\partial}{\partial R}\left(R\vr\siggas R^2\Omega\right)- 
\frac{1}{2\pi}\frac{\partial{\calg}}{\partial{R}}+R^3\Omega\sigsfr=0, 
\end{equation} 
where $\calg=2\pi R^3\nu\siggas\left(d\Omega/dR\right)$ is the viscosity torque, $\nu$ is kinematic 
viscosity, and $\Omega$ is the angular velocity. Combining eqs. (1) and (2), we have 
\begin{equation} 
\frac{\partial{\siggas}}{\partial{t}}=-\frac{1}{2\pi R}\frac{\partial{}}{\partial{R}} 
     \left\{\left[\frac{d \left(R^2\Omega\right)}{dR}\right]^{-1} 
     \frac{\partial \calg}{\partial R}\right\}-\sigsfr. 
\end{equation} 
In this Letter, we explore the role of viscosity driven by SNexp in the secular evolution of 
gaseous disk. Similar to Kawakatu \& Wada (2008), we assume that the vertical structure 
is supported by turbulence pressure  
\begin{equation} 
\frac{P_{\rm tur}}{H}=\rho_{\rm g}\left(\frac{G\mbh H}{R^3}+\pi G\Sigma\right) 
                     \approx \rho_{\rm g}\frac{\sigma^2}{R}, 
\end{equation} 
where $\mbh$ is the black hole mass, $G$ is the gravity constant, $\Omega=\sqrt{2}\sigma/R$ and
$\pi G\Sigma=\sigma^2/R$ are used
(eq. 1 in Thompson et al. 2005), $P_{\rm tur}=\rho_{\rm g}V_{\rm tur}^2$ is the turbulence pressure, 
$\sigma$ is the dispersion velocity of galaxies determined by the total potential of gas, stars, and 
dark matter halo, and $H$ is the half-thickness of the disk. The validity of solutions presented here is 
limited by the potential used in eq. (4), namely, $G\mbh H/R^3\lesssim \sigma^2/R$, we thus have,  
$R\gtrsim G\mbh(H/R)/\sigma^2\sim 10~M_8\sigma_{200}^2(H/R){\rm pc}$, where $M_8=\mbh/10^8M_\odot$ and  
$\sigma_{200}=\sigma/200~{\rm km~s^{-1}}$.  
The energy equation of turbulence excited by SNexp is given by Wada \& Norman (2002) 
\begin{equation} 
\frac{\rho_{\rm g}V_{\rm tur}^2}{t_{\rm dis}}=\frac{\rho_{\rm g}V_{\rm tur}^3}{H} 
                                             =\xi {\dot S}_{\star}E_{\rm SN}, 
\end{equation} 
where $t_{\rm dis}=H/V_{\rm tur}$ is the timescale of the turbulence, 
$\xi$ is the efficiency converting kinetic energy of SNexp into turbulence, 
$\dot{S}_\star$ is the SNexp rate, and $E_{\rm SN}$ is the SNexp energy. 
SNexp rate strongly depends on the initial mass functions, but
we absorb its uncertainties into the parameter $\xi$ and simply let $\dot{S}_\star$ 
be the star formation rate.

The viscosity law $\nu=\alpha V_{\rm tur}H$ is assumed commonly, where $\alpha$ is the viscosity 
parameter (Shakura \& Sunyaev 1973). The Kennicutt--Schmidt's law displays 
$\dot{\Sigma}_\star\propto \Sigma^{1.4}$, however, there are growing evidence for a linear relation 
for denser environment in star-forming galaxies from numerical simulations (Dobbs \& Pringle 2009; 
Krumholz et al. 2009). We use $\sigsfr=c_{\star}\siggas$ for star-forming galaxies at high-redshift, 
where $c_\star$ is a constant. We thus have the energy equation  
$V_{\rm tur}^3 =\xi c_\star H \esn$, where $c_\star$ is the efficiency converting gas into stars.  
With the help of equations (4) and (5), we have 
\begin{equation} 
\nu(R)=\zeta R^4=0.128 ~\zeta_{0.128}R_1^4~{\rm kpc^2~Gyr^{-1}}, 
\end{equation} 
where $\zeta=\alpha\left(\xi c_\star \esn\right)^3/\sigma^8 
            =0.128\alpha_{0.1}\left(\xi_{-3}c_{-8}E_{51}\right)^3\sigma_{200}^{-8}~$ 
	    ${\rm kpc^{-2}Gyr^{-1}}$,   
	    $R_{1}=R/1~{\rm kpc}$, $\alpha_{0.1}=\alpha/0.1$, $E_{51}=E_{\rm SN}/10^{51}{\rm erg}$, 
	    $\xi_{-3}=\xi/10^{-3}\sunm^{-1}$, and $c_{-8}=c_*/3\times 10^{-8}~{\rm yr^{-1}}$. 
Here, the star formation efficiency is assumed to be a constant independent of the gas density and 
radius. Such strong dependence of viscosity on the radius means that transportation of angular 
momentum will be much more efficient at outer radii than at the inner.  

The star surface density can be obtained from  
\begin{equation} 
\Sigma_*(R,t)=\int_0^t \sigsfr d\tp=\int_0^t c_{\star}\siggas (R,\tp) d\tp. 
\end{equation} 

Vertical self-gravity may directly control the star formation as suggested by Thompson et al. 
(2005). We test self-gravity of the disk through Toomre parameter $Q$, defined by 
$Q=c_s\kappa_{\tiny\Omega}/\pi G \siggas\propto \siggas^{-1}$, where 
$\kappa_{\tiny\Omega}^2=4\Omega^2+d\Omega^2/d\ln r$ is the epicyclic frequency, 
$c_s=1.0T_2^{1/2}~{\rm km~s^{-1}}$is the sound speed, and $T_2=T/10^2 {\rm K}$ is the gas 
temperature (Kawakatu \& Wada 2008).  We should keep in mind that there are 
two different aspects from the classical gaseous disk investigated by Lynden-Bell \& Pringle 
(1974): (1) radius-dependent viscosity driven by SNexp and (2) dropout of mass and angular momentum 
due to star formation on the disk. These drive some interesting properties of gaseous disks.

\section{Solutions: evolutionary properties} 
\subsection{Gaseous ring} 
A realistic initial condition is a single gaseous ring, which can be formed via one minor merger 
(Hernquist \& Mihos 1995). Equations (3-6) have analytical solutions for the case of a single ring. 
After some algebraic operations (e.g., Kato et al. 1998), we have the analytical solution for the case 
\begin{equation} 
\siggas(r,\tau)=\frac{\siggas_0}{2\pi^{\frac{1}{2}}}~\frac{1}{r^4}
                \left(\frac{q_0}{\tau}\right)^{\frac{1}{2}} 
	        \left(1-e^{-\frac{4q_0}{\tau r}}\right) 
                e^{-\tau-\frac{q_0}{\tau}\left(\frac{1}{r}-1\right)^2}, 
\end{equation} 

{\centering 
\figurenum{2} 
\includegraphics[angle=-90,scale=0.55]{fig2.ps} 
\figcaption{\footnotesize Properties of the surface density $\Sigma(r,\tau)$ as to $q_0$. Here, $\rp$ 
and $\taup$ are solutions of $\partial \Sigma /\partial r=0$ and $\partial \Sigma/\partial \tau=0$. 
$\Sigma(\rp, \taup)$ is the peak or valley values at $\rp$ and $\taup$. The red and blue points 
correspond to peak and valley of the density, respectively. The stars represent $r_{\rm p}$. When 
$q_0>q_{\rm c}=0.544$, there is no solution, namely, the density monotonously decreases with time 
and radius. All the hollow symbols are the valley of the envelope of the $\siggas$-evolution track 
whereas the solid ones are its peak.  
} 
\label{fig2} 
\vglue 0.3cm}

where $\siggas_0=M_0/\pi R_0^2$, $\tau=c_*t$, $r=R/R_0$, and $M_0$ is the total mass  
of the initial ring at a radius $R_0$, and the dimensionless parameter $q_0$ is defined by 
\begin{equation} 
q_0=\frac{c_*}{4\zeta R_0^2}=0.584~\alpha_{0.1}^{-1}c_{-8}^{-2}R_{0,1}^{-2} 
                           \left(\xi_{-3}E_{51}\right)^{-3}\sigma_{200}^8, 
\end{equation} 
where $R_{0,1}=R_0/10~{\rm kpc}$.

Complicated evolutionary behaviors of the surface density are determined by $q_0$ values as shown in 
Figure 2. For a given $q_0\le q_c=0.544$, there are two sets of roots of 
$\partial \Sigma(r,\tau)/\partial r=0$ and $\partial \Sigma(r,\tau)/\partial \tau=0$, corresponding 
to the peak and valley of the envelope 

of $\Sigma$-evolution track at $r_{\rm p}$ and $\tau_{\rm p}$. 
Otherwise there is no root of the two equations, namely there is no peak or valley of the envelope. 
Evolution of $\siggas(r,\tau)$ and $\Sigma_*(r,\tau)$ can be found in Figure 3(a) and (b), 
respectively. We show three types of the solutions: (1) $q_0< q_c$, (2) $q_0=q_c$, and (3) $q_0>q_c$. 
The $q_0>q_c$ solutions are quite simple and similar to the pure diffusion of a gaseous ring described 
in Lynden-Bell \& Pringle (1974). The ring dramatically spreads over the space and forms stars. A 
stellar ring forms then as shown in Figure 3(b) $q_0=2$ panel. For $q_0=q_c$ solution, there 
is a flat envelope without peak and valley, forming a very broader stellar ring as shown in the 
Figure 3(b) $q_0=0.544$ panel. 

The $q_0<q_c$ solutions have complicated behaviors as shown in Figure 3(a) $q_0=0.1$ panel. Evolution 
of the gas ring can be divided into three phases. At early time, star formation is only important in the 
role of transportation of angular momentum, and the mass dropout can be neglected. Behaviors of the ring 
are very similar to the pure fluid ring. The gas rapidly diffuses inward mostly and outward a little bit 
carrying away the angular momentum. Though the star formation rates are very high, the total formed
stars are not many since the duration of this phase is quite short. The second phase begins when mass 
dropout due to star formation becomes important. Decrease of the gas density is leading to decrease of 
star formation and weakening the viscosity torque of SNexp, resulting in accumulation of gas at the radius 
$r_{\rm p}$. The third phase starts when the viscosity torque is enhanced in light of the gas accumulation. 
This causes intensive star formation and exhausts gas, giving rise to an exponential decreases of gas 
density.

{\centering
\figurenum{3} 
\includegraphics[angle=-90,scale=0.4]{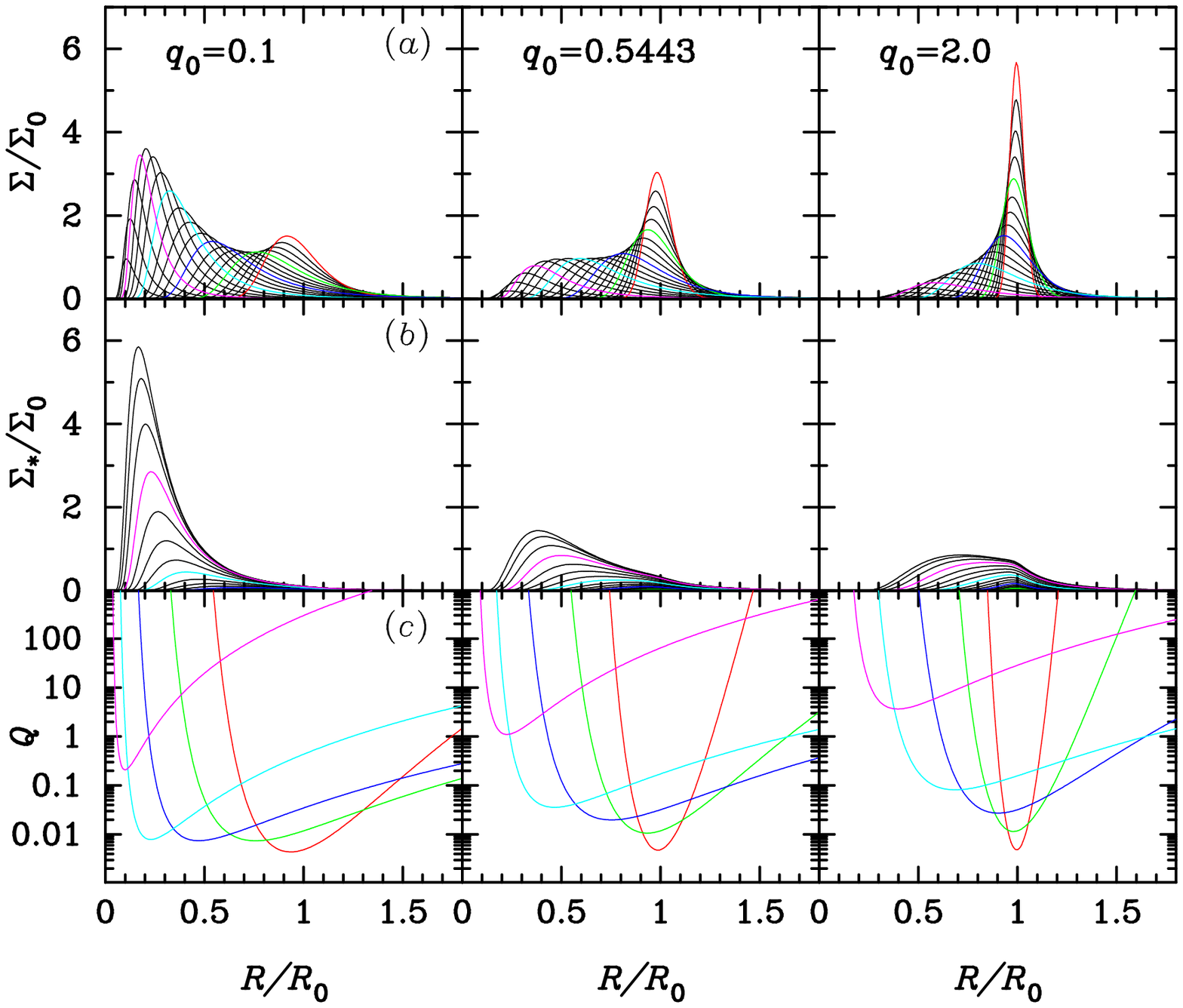} 
\figcaption{\footnotesize Evolution of gaseous ring and star surface densities for different parameter 
$q_0$. In top and middle panel, black lines show the evolution of gas and star disk from $\tau_1=0.005$ 
to $\tau_2=5$ with an interval $\Delta \log \tau=0.15$.
Red, green, blue, cyan, and pink lines in bottom panel represent the Toomre parameter $Q$ at $\tau=0.005,
0.03, 0.2, 1, and 5$, respectively, which correspond to the same meanings in panels (a) and (b). 
We set the surface density of gas is $\Sigma_0=100\sunm~{\rm pc^{-2}}$ 
to calculate the Toomre parameter for three different $q_0$ values.
} 
\label{fig3} 
\vglue 0.3cm}

To understand such behaviors of solutions, we compare the timescales of star formation ($t_*=1/c_*$) 
and gas advection ($t_{\rm adv}=R/V_R\approx R^2/\nu=1/\zeta R^2$). According to the $q_0$-definition, 
it follows $t_{\rm adv}/t_*=4q_0$. When $t_*\ll t_{\rm adv}$ at $R_0$ (roughly corresponding to 
$q_0\gg q_c$), most of the gas is converted into stars at $R_0$ and only a little gas is conveyed 
inward. In contrast, in the $t_*\gg t_{\rm adv}$ case (namely, $q_0\ll q_c$) at initial radius $R_0$, 
star formation is so slow that most of the gas is advected inward. For a case with initial $q_0\ll q_c$, 
most of gas is advected inward until $t_{\rm adv}>t_*$. Pile-up of gas appears at $r_{\rm p}$ then 
and forms stars.  The complicated behaviors of gas density are results as a consequence of competition 
between the two processes.

Figure 3(b) shows the surface density of stars for the corresponding gas density. For $q_0=0.1$, the 
gas rapidly diffuses inward and there is no significant pile-up of stars at the initial radius $R_0$. 
A stellar ring forms at roughly $\sim 0.15 R_0$ far away from its initial radius. The ring becomes 
broader with increases of $q_0$. When $q_0=q_c$, there begins to be a tiny pile-up of stars at the 
initial radius. For a case with a large $q_0$, pile-up of stars happens very close to the initial 
radius and it becomes more conspicuous with $q_0$. The stellar rings become wider with $q_0$.
Figure 3(c) displays evolution of the Toomre parameter $Q$, clearly showing $Q<1$ in the star-forming
region. $Q$-values are quite different in different regions. This strengthens the importance 
of self-gravity in the secular evolution of the gaseous disk. Since $Q\propto \Sigma_{\rm gas}^{-1}$, 
$Q$ basically follows the evolution of $\Sigma_{\rm gas}$. A generic property of the $Q$-parameter 
is that it increases with time since the gas is being converted into stars. This is totally different 
from the steady gaseous disk with star formation under the presumed condition of $Q=1$ suggested by 
Thompson et al. (2005).   

{\centering
\figurenum{4} 
\includegraphics[angle=-90,scale=0.4]{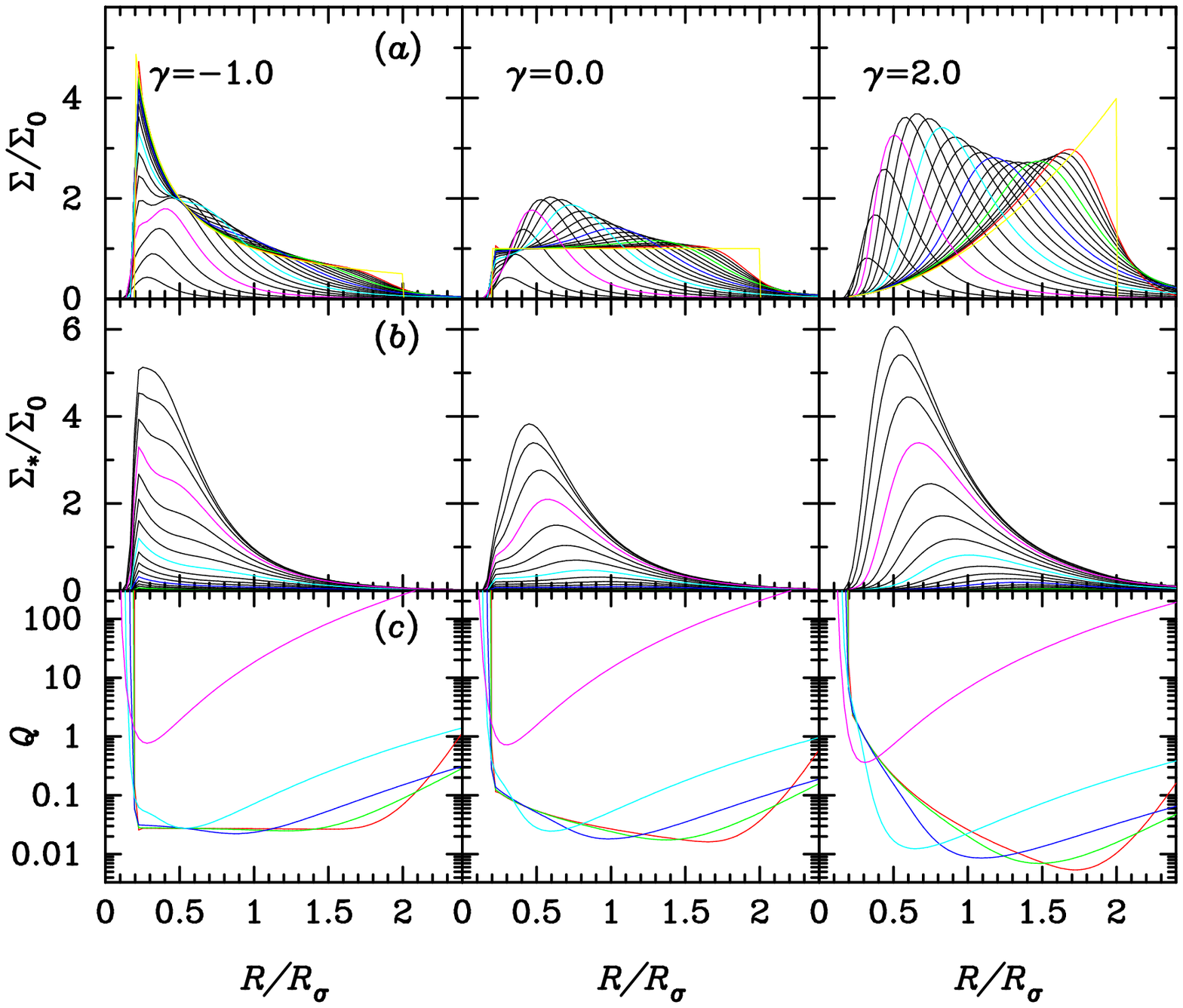} 
\figcaption{\footnotesize Evolution of gaseous disk and star surface densities for different initial disks. 
In the top and middle panel, black lines show the evolution of gas and star disk from $\tau_1=0.005$ to 
$\tau_2=5$ with an interval $\Delta \log \tau=0.15$. Red, green, blue, cyan, and pink lines in bottom panel 
represent the Toomre parameter $Q$ at $\tau=0.005, 0.03, 0.2, 1, and 5$, respectively, which correspond to the
same meanings with panels (a) and (b). We set the surface density 
of gas is $\Sigma_0=100\sunm~{\rm pc^{-2}}$ to calculate the Toomre parameter for three different $\gamma$ 
values. The yellow lines in (a) are the initial surface densities.
} 
\label{fig4}
\vglue 0.3cm 
}

We would like to point out that the feedback from SNexp will be enhanced, if we use the non-linear 
Kennicutt--Schmidt's law to calculate the star formation rate, leading to more efficient viscosity
to transport angular momentum. However, the main features given by eq. (8) will remain. A detailed
discussion on this nonlinear effects is beyond the scope of this Letter, but it would be 
interesting to solve the nonlinear equations to display the effects in future work. This model 
may apply to different kinds of galaxies.
 
\subsection{Gaseous disk} 
Successive multiple minor mergers may provide an initial gas widely distributed in the major galaxies 
(Hernquist \& Mihos 1995; Bournaud et al. 2007). For a simplicity, we assume an initial distribution
with a power law of radius as $\siggas(R,0)=\Sigma_0 \left(R/\Rsigma\right)^{\gamma}$, where
$\Rsigma=\left(c_*/4\zeta\right)^{1/2}=7.65~c_{-8}^{1/2}\zeta_{0.128}^{-1/2}~$kpc, $\Sigma_0$ is 
the surface density at $\Rsigma$, and $\gamma$ is the index. 
We integrate the Green's function over the radius range of the disk, and have 
\begin{equation} 
\siggas(\rsigma,\tau)=\int_{0}^{\infty}\frac{\siggas(r_0,0)}{(\pi \tau)^{\frac{1}{2}}} 
                      \frac{r_0^2}{\rsigma^4}\left(1-e^{-\frac{2r_0^{-1}}{\tau\rsigma}}\right) 
		      e^{-\tau-\frac{1}{\tau}\left(\frac{1}{\rsigma}-\frac{1}{r_0}\right)^2} 
                      d r_0,
\end{equation} 
where $\rsigma=R/\Rsigma$ and $r_0=R_0/\Rsigma$. 
  
Figure 4 shows the evolution of gaseous disk. 
We calculate the cases of $\gamma=-1.0,0, and 2.0$ between $R_1=0.2\Rsigma$ and $R_2=2.0\Rsigma$.
For $\gamma=-1.0$ disk, the most dense part of the disk is located in the radius with
$t_*<t_{\rm adv}$, so the gas is converted to stars locally. In contrast, for $\gamma=2.0$
case gas is concentrated outside the region with $t_*>t_{\rm adv}$. The gas diffuses quite rapidly 
and transfers to the inner radius and forms stars, leaving a relatively broader stellar ring or disk.
An initial homogeneous disk has a complex behaviors between the two cases above, inner part of the gas 
forms stars locally whereas outer part flows inward and then forms stars. The Toomre parameter shows 
strong variability with time, but holding $Q<1$ in most star-forming region.

{\centering
\figurenum{5} 
\includegraphics[angle=-90,scale=0.35]{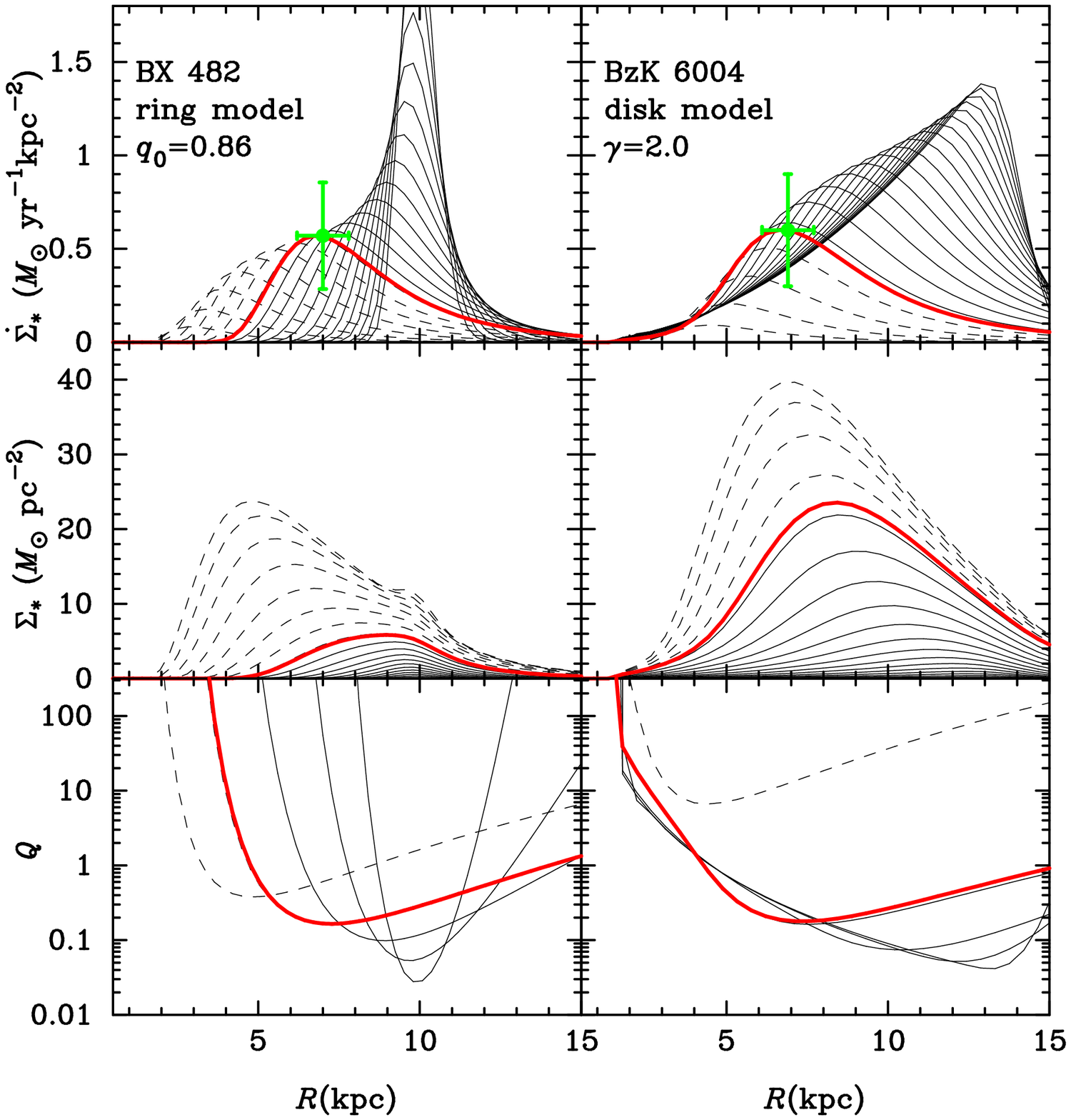} 
\figcaption{\footnotesize All the red thick lines are the current status of galaxy BX 482 and BzK 6004. 
Given $\sigma=210\kms$ from observations for the model parameters of BX 482, we relatively arbitrarily 
choose $R_0=10$kpc, then make out $q_0=0.86$. We have $\Sigma_0=9.3\sunm~{\rm pc^{-2}}$ from star 
formation rates. For Bzk 6004, we find that the initial parameters are $\Sigma_0=69\sunm~{\rm pc^{-2}}$,
$R_1=1.6, R_2=14.2$kpc, $\gamma=2$, and $t=3.3\times 10^7$yr. The solid lines represent the past status 
of the two galaxies and dotted lines do there future status. 
} 
\label{fig5}
\vglue 0.3cm 
}

The instability of the gaseous disks can be simply justified by the operating mechanism of the 
viscosity. When gas is piled up, the higher star formation rates can efficiently remove the angular 
momentum of gas, and then suppress the star formation rates. The system should be self-organized and 
stable. However, we should give a detail description of stability analysis in a future paper.

\section{Applications}
Dynamics of the gaseous disk presented here allows us to compare current model with Genzel et al.'s 
observations of star-forming galaxies at high-redshift. Observations of SINFONI on ESO VLT provide 
details of the "cold gas" accretion. We choose two galaxies as illustrations to apply the present 
theoretical model. BX 482 is a ring-dominated star-forming galaxy, of which the ring radius is 
$R_{\rm ring}=7.0\pm 0.8$kpc with a narrower width $\Delta R\sim 1$ kpc, its dispersion velocity 
is $\sigma=210$km~s$^{-1}$ and the surface density of star formation rates is 
$0.57\pm 0.3\sunm~{\rm yr^{-1}~kpc^{-2}}$. We hence choose the ring model (eq. 8). The bright 
H$\alpha$ ring limits the lifetime of the ring, which is about $\sim 10^7$yr in light of OB stars. 
The left column of Figure 5 gives the dynamical evolution of the ring. The Toomre parameter $Q$ shows 
that the ring is self-gravity dominated. A stellar ring formed in the gaseous ring is still growing 
shown by the red line, but the H$\alpha$ ring is becoming dimmer. 

BzK 6004 is also a rotation-dominated galaxy with relatively faint but broader H$\alpha$ ring 
($\Delta R/R\sim 1.0$ from Figure 5 in Genzel et al. 2008). The ring radius is $R_{\rm ring}=6.9\pm 0.8$kpc 
and the dispersion velocity of galaxy is $\sigma=240\kms$ and the surface density of star formation rate 
is $0.6\pm 0.3\sunm~{\rm yr^{-1}~kpc^{-2}}$. The current data does not allow us to distinguish the case of 
different $\gamma$ we thus just use $\gamma=2$ as an illustration. The dynamical evolution of the galaxy 
is shown by the red line in the right column of Fig. 5. The current stage of the ring is approaching the 
end states of the star forming. The stellar ring is still growing fast and may remain as a wide stellar 
ring. The $Q$-value shows that self-gravity is also dominated. 

We note that the solutions presented here tend to show an exponential stellar disk if the time is long 
enough. This agrees with the steady disks known for quite long time from Lin \& Pringle (1987).
The simple application of the current model shows here that SNexp does indeed play a key role in the 
dynamical evolution of the gaseous rings or disks. However, the application should be improved with 
the more detail data. For example, the good-quality data of H$\alpha$ ring profile and the spatially 
resolved spectroscopy will allow us to get the evolution stage of gas and star formation from the 
stellar synthesis. Particularly, the very sharp inner edge of the ring predicted here is caused by 
SNexp. This reflects feedback of the star formation. Additionally, application of the present model 
to a large sample (e.g. F\"oster Schreiber et al. 2009) will produce the initial conditions of the 
sample for a statistics in future. This will be invaluable to justify the origin of the cold gas.

We would like to stress here that the present applications focus on the main features of galaxies 
at high-redshift. The ring or disk models for the two galaxies are less conclusive. We need more  
information of the galaxies, such as, stellar ages at different ring radii, so as to give more 
robust pictures. Degeneracies of parameters limit applications of the model.

\section{Conclusions} 
We set up a model of gaseous disk with star formation, in which the viscosity is driven by turbulence 
excited by SNexp. We get the analytical solutions for the initial condition of single  
gas ring, showing complicated secular evolutionary behaviors. 
Several different kinds of stellar distributions can be formed, including exponential types.  
The present model is able to reproduce starburst rings in two galaxies as illustrations. 
 
The model will be improved in several aspects in future by using: (1) nonlinear star formation law,
(2) separated equations of stars and gas, and (3) initially twisted gaseous disk (Lu \& Cheng 1991). These 
will help us to get the initial conditions as a consequence of minor mergers and allow us to study
the evolution of stellar disks and bulges.

\acknowledgements{The referee is greatly thanked for a very helpful report improving the manuscript.
Prof. J.-F. Lu is acknowledged for useful discussions.
We appreciate the stimulating discussions among the members of IHEP AGN  
group. The research is supported by NSFC-10733010 and 10821061, CAS-KJCX2-YW-T03, and
973 project (2009CB824800). }

\end{document}